\documentclass[twocolumn]{aastex62}
\usepackage{epstopdf}
\usepackage{microtype}  
\usepackage{booktabs}   

\received{08/09/2018}
\revised{09/13/2018}
\accepted{09/14/2018}
\submitjournal{ApJ Letters}
\shorttitle{\emph{NICER} Low-Freq.~QPO in MAXI J1535--571}
\shortauthors{Stevens et al.}

\newcommand{\chisq}{\ensuremath{\chi^{2}}}
\newcommand{\vcent}{\ensuremath{\nu_{\text{centroid}}}}
\newcommand{\degrees}{\ensuremath{^{\circ}}}

\begin{document}
\title[\emph{NICER} Low-Frequency QPO]{A \emph{NICER} Discovery of a Low-Frequency Quasi-Periodic Oscillation in the Soft-Intermediate State of MAXI~J1535--571}

\author[0000-0002-5041-3079]{A.~L.~Stevens}
\affiliation{Department of Physics \& Astronomy, Michigan State University, 567 Wilson Road, East Lansing, MI 48824, USA}
\affiliation{Anton Pannekoek Institute for Astronomy, University of Amsterdam, Science Park 904, 1098 XH Amsterdam, The Netherlands}
\affiliation{Department of Astronomy, University of Michigan, 1085 South University Avenue, Ann Arbor, MI 48109, USA}
\email{alstev@msu.edu}

\author{P.~Uttley}
\affiliation{Anton Pannekoek Institute for Astronomy, University of Amsterdam, Science Park 904, 1098 XH Amsterdam, The Netherlands}

\author[0000-0002-3422-0074]{D.~Altamirano}
\affiliation{Physics and Astronomy, University of Southampton, Southampton SO17 1BJ, UK}

\author{Z.~Arzoumanian}
\affiliation{Astrophysics Science Division, NASA Goddard Space Flight Center, Greenbelt, MD 20771, USA}

\author{P.~Bult}
\affiliation{Astrophysics Science Division, NASA Goddard Space Flight Center, Greenbelt, MD 20771, USA}

\author[0000-0002-8294-9281]{E.~M.~Cackett}
\affiliation{Department of Physics \& Astronomy, Wayne State University, 666 West Hancock Street, Detroit, MI 48201, USA}

\author{A.~C.~Fabian}
\affiliation{Institute of Astronomy, University of Cambridge, Madingley Road, Cambridge CB3 0HA, UK}

\author{K.~C.~Gendreau}
\affiliation{Astrophysics Science Division, NASA Goddard Space Flight Center, Greenbelt, MD 20771, USA}

\author{K.~Q.~Ha}
\affiliation{Applied Engineering and Technology Directorate, NASA Goddard Space Flight Center, Greenbelt, MD 20771, USA}

\author[0000-0001-8371-2713]{J.~Homan}
\affiliation{Eureka Scientific, Inc., 2452 Delmer Street, Oakland, CA 94602, USA}
\affiliation{SRON, Netherlands Institute for Space Research, Sorbonnelaan 2, 3584 CA Utrecht, The Netherlands}

\author{A.~R.~Ingram}
\affiliation{Department of Physics, Astrophysics, University of Oxford, Denys Wilkinson Building, Keble Road, Oxford OX1 3RH, UK}

\author[0000-0003-0172-0854]{E.~Kara}
\affiliation{Department of Astronomy, University of Maryland, College Park, Maryland 20742, USA}

\author{J.~Kellogg}
\affiliation{Applied Engineering and Technology Directorate, NASA Goddard Space Flight Center, Greenbelt, MD 20771, USA}

\author[0000-0002-8961-939X]{R.~M.~Ludlam}
\affiliation{Department of Astronomy, University of Michigan, 1085 South University Avenue, Ann Arbor, MI 48109, USA}

\author{J.~M.~Miller}
\affiliation{Department of Astronomy, University of Michigan, 1085 South University Avenue, Ann Arbor, MI 48109, USA}

\author[0000-0002-8247-786X]{J.~Neilsen}
\affiliation{Villanova University, Mendel Hall, Room 263A, 800 East Lancaster Avenue, Villanova, PA 19085, USA}

\author[0000-0003-1386-7861]{D.~R.~Pasham}
\affiliation{MIT Kavli Institute for Astrophysics and Space Research, MIT, 70 Vassar Street, Cambridge, MA 02139, USA}

\author[0000-0003-4815-0481]{R.~A.~Remillard}
\affiliation{MIT Kavli Institute for Astrophysics and Space Research, MIT, 70 Vassar Street, Cambridge, MA 02139, USA}

\author[0000-0002-5872-6061]{J.~F.~Steiner}
\affiliation{MIT Kavli Institute for Astrophysics and Space Research, MIT, 70 Vassar Street, Cambridge, MA 02139, USA}

\author[0000-0002-5686-0611]{J.~van~den~Eijnden}
\affiliation{Anton Pannekoek Institute for Astronomy, University of Amsterdam, Science Park 904, 1098 XH Amsterdam, The Netherlands}

\begin{abstract}

We present the discovery of a low-frequency $\approx$\,5.7\,Hz quasi-periodic oscillation (QPO) feature in observations of the black hole X-ray binary MAXI~J1535--571 in its soft-intermediate state, obtained in 2017 September--October by the \emph{Neutron Star Interior Composition Explorer} (\emph{NICER}).  
The feature is relatively broad (compared to other low-frequency QPOs; quality factor $Q\approx 2$) and weak (1.9\% rms in 3--10\,keV), and is accompanied by a weak harmonic and low-amplitude broadband noise.  
These characteristics identify it as a weak Type A/B QPO, similar to ones previously identified in the soft-intermediate state of the transient black hole X-ray binary XTE~J1550--564.  
The lag-energy spectrum of the QPO shows increasing soft lags toward lower energies, approaching 50\,ms at 1 keV (with respect to a 3--10\,keV continuum). 
This large phase shift has similar amplitude but opposite sign to that seen in \emph{Rossi X-Ray Timing Explorer} data for a Type~B QPO from the transient black hole X-ray binary GX~339--4.  
Previous phase-resolved spectroscopy analysis of the Type~B QPO in GX~339--4 pointed toward a precessing jet-like corona illuminating the accretion disk as the origin of the QPO signal.
We suggest that this QPO in MAXI~J1535--571 may have the same origin, with the different lag sign depending on the scale height of the emitting region and the observer inclination angle.

\end{abstract}
\keywords{accretion, accretion discs --
black hole physics --
stars: black holes --
X-rays: binaries -- 
X-rays: individual: MAXI~J1535--571}

\section{Introduction} \label{sec:intro}

MAXI~J1535--571 is a newly discovered transient X-ray binary that went into outburst starting on 2017 September 2.
It was independently discovered in the X-rays by the \emph{Monitor of All-sky X-Ray Image} (\emph{MAXI}; \citealt{Negoroetal17a}) and the \emph{Neil Gehrels Swift Observatory} \citep{Markwardtetal17,Kenneaetal17a}.
Its multi-wavelength properties strongly suggested that the source is an accreting stellar-mass black hole, initially seen in the hard spectral state \citep{Scaringietal17a,Scaringietal17b,Negoroetal17b,Russelletal17a,Dincer17,Brittetal17,Xuetal18}.

One week following its discovery, X-ray, radio, and sub-mm detections showed the source brightening and entering the hard-intermediate spectral state \citep{Nakahiraetal17a,Kennea17b,Palmeretal17,Tetarenkoetal17,Shidatsuetal17a}.
In the hard-intermediate state, low-frequency quasi-periodic oscillations (LF QPOs) were detected in the X-ray light curve by \emph{Swift} \citep{Mereminskiy17,Russelletal17b} and the \emph{Neutron Star Interior Composition Explorer} (\emph{NICER}; \citealt{Gendreauetal17}). 
Detailed spectral fits to a \emph{NICER} observation from 2017 September 13 showed a narrow Fe~K emission line and strong reflection features, pointing toward a nearly-maximal black hole spin and a possible warp in the accretion disk \citep{Milleretal18}.

The source continued to transition into the soft state in 2017 October and November \citep{Shidatsuetal17b}.  
Although heavily extinguished (neutral column density $N_{\rm H}\sim$~few$\times 10^{22}$~cm$^{-2}$), MAXI~J1535--571 reached extremely bright flux levels (up to 5~Crab in 2--20\,keV flux; \citealt{Shidatsuetal17a}).
The source faded significantly in the soft state through the end of 2018 April to a 2--10\,keV flux of $<$\,8\,mCrab \citep{Negoroetal18}, then showed repeated transitions between soft and hard spectral states at these low flux levels, beginning in early 2018 May \citep{Russelletal18,Parikhetal18} and presently ongoing.\footnote{For up-to-date information on MAXI~J1535--571 from \emph{MAXI}, see \url{http://maxi.riken.jp/nakahira/1535monitor/}.}

LF QPOs are associated with Fourier frequencies of $\sim$\,0.1 to tens of Hz in the X-ray light curves of accreting black hole X-ray binaries. 
The three types of LF QPOs seen in black hole X-ray binaries are classified as Types A, B, and C \citep{Wijnandsetal99,Remillardetal02,Casellaetal05,Mottaetal11}. 
Over an entire outburst, Type C QPOs are the most commonly seen, Type B are sometimes seen, and Type A are rarely seen.
Type A LF QPOs have low-amplitude broadband noise and no harmonic in the average power spectrum, and they appear in the soft-intermediate state of an outburst when the light curve has rms\,$\lesssim$\,3\% in the 2--60\,keV energy band. 
Type B LF QPOs are characterized by weak harmonics and weak broadband noise compared to the QPO amplitude, and appear in the soft-intermediate state when the rms of the light curve is $\approx$\,3\%--5\% in 2--60\,keV (see, e.g., \citealt{Heiletal15a}).
Type C LF QPOs are characterized by strong harmonics and strong broadband noise, and appear in the hard-intermediate state when the rms of the light curve is $\approx$\,5\%--30\% in 2--60\,keV. 
Type A QPOs have a weak, broad QPO peak at $\approx$\,8\,Hz, Type B QPOs generally have a strong peak at $\lesssim$\,6\,Hz (\citealt{Mottaetal11} found that the Type B QPO peak frequency was anti-correlated with the source flux in GX~339--4), 
and Type C QPOs have peaks that tend to move or shift in the frequency range $\sim$\,0.1--20\,Hz in (anti-)correlation with evolving spectral parameters (see, e.g., \citealt{Vignarcaetal03,Stieleetal13}).
The relative width of the QPO can be quantified by the coherence, $Q$, computed as a ratio of the QPO's centroid frequency (\vcent) to its FWHM; Type A typically have $Q\lesssim 3$, Type B typically have $Q\gtrsim 6$, and Type C typically have $7\lesssim Q\lesssim 12$ \citep{Casellaetal05}.

Theories of the origin of LF QPO variability encompass several physical explanations that can broadly be categorized as geometric or intrinsic quasi-periodic variability.
Systematic observational studies suggest that the LF QPO origin is linked to the geometry of the system due to the correlation of the QPO amplitude \citep{Schnittmanetal06,Mottaetal15,Heiletal15b} and the sign of the energy-dependent lags \citep{vdEijndenetal17} with the binary orbit inclination.  
Type B and C QPOs show different amplitude-inclination correlations \citep{Mottaetal15}, which may suggest distinct origins. 

Spectral-timing analysis reveals the nature and scaling of causal relationships between variations in the distinct emission regions in accreting black hole systems.  
These components can be identified from the X-ray spectrum (e.g., disk blackbody and Comptonized emission).  
Lags may be computed from the phase of the Fourier cross spectrum to measure the energy-dependent delays of variability in a narrow energy band of interest with respect to a (broad) reference band \citep{Vaughanetal94,VaughanNowak97}, and relate them to the components seen in the X-ray spectrum.
For a review and description of some of the key spectral-timing techniques applied to broadband noise, see \citet{Uttleyetal14}. 

In this Letter, we present the power and lag-energy spectral analysis of a weak Type A/B QPO from MAXI~J1535--571 detected early in its outburst during a soft-intermediate state.  
We present spectral-timing data from the \emph{NICER} payload operating on the International Space Station since 2017 June.
\emph{NICER}'s large collecting area, soft X-ray response, $\approx$\,100\,eV energy resolution, $\approx$\,100\,ns time resolution, and high count-rate capability enable unprecedented spectral-timing studies of QPOs, even of the weak feature we report here.
In Section \ref{sec:data}, we describe the data and data reduction procedure. 
In Section \ref{sec:results}, we present the data selection criteria, average power spectrum, and lag-energy spectrum.
The discussion and conclusions are presented in Section \ref{sec:discuss}.

\section{Data} \label{sec:data}

The \emph{NICER} X-Ray Timing Instrument (XTI) delivers photons onto 56 focal plane modules (FPMs) that are grouped into seven sets of eight, with each set read out by one measurement/power unit (MPU; \citealt{Gendreauetal12,Arzoumanianetal14}).  
FPMs 11, 20, 22, and 60 are not operational, so they do not contribute events to the \emph{NICER} dataset.
Furthermore, FPMs 14 and 34 were removed from our analysis.\footnote{FPM 14 was removed because it has been found to intermittently exhibit faults in its processing that can result in spurious energy spectral features (K.~Hamaguchi 2017, private communication). FPM 34 is exceptionally sensitive to a solar light-leak that results in excessive noise, to the extent that noise contribution from this single FPM can exceed the rest of the XTI ensemble.}
Our ``good'' selection consists of 50 operational FPMs, providing 
an effective area of approximately 1800\,cm$^{2}$ at 1.5\,keV.

In this Letter we present the analysis of \emph{NICER} XTI observations obtained in 2017 September and October.  
These data correspond to ObsIDs 1050360101 through 1050361020 and 1130360101 through 1130360114. 
With a near-daily sampling of one or multiple few-100\,s exposures, these observations provide a total exposure of $\approx$\,122\,ks and $\approx$\,$1.6\times10^{9}$ photon counts.
Based on the source spectral and timing properties (see Section \ref{sec:lfqpo}), these observations correspond to the hard-intermediate and soft-intermediate states of accreting black holes.

The data were processed with \textsc{nicerdas} version 2018-03-01$\_$V003.
The data were cleaned using standard calibration with \textsc{nicercal} and standard screening with \textsc{nimaketime}. 
We selected events that were detected outside the South Atlantic Anomaly, more than 40\arcdeg{} away from the bright Earth limb, more that 30\arcdeg{} away from the dark Earth limb, less than 54\arcsec{} offset in pointing, not flagged as ``overshoot'' or ``undershoot'' resets, and triggered the slow signal chain (\texttt{EVENT\_FLAGS = x1x000}). We also applied a ``trumpet filtering'' to remove known background events.
These processing steps were automatically carried out in the standard \emph{NICER} data reduction pipeline.

For our analysis, we split the \emph{NICER} event list light curves into uninterrupted sequential 64\,s segments.
The XTI has an overall photon time-stamping resolution of 85\,ns (1$\sigma$), but we bin light curves to a time step of $2^{-8}$\,s ($\simeq$\,3.9\,ms). 
For 64\,s light curve segments, this gives 16,384 time bins apiece.
This time binning gives a Nyquist frequency of 128\,Hz; having a Nyquist frequency much higher than our signal frequency lets us check for deadtime effects on the Poisson noise level (see Section \ref{sec:discuss}).

Though the background contribution is expected to be negligible ($\sim$\,2\,count\,s$^{-1}$; \citealt{Keeketal18, Ludlametal18}),
we applied a filter to remove segments that showed signs of flaring or non-source particle background. 
We calculated the average count rate in each FPM for each segment, and if any FPM count rate was more than four times half the interquartile range away from the median count rate of all FPMs, that segment was discarded.
Out of 1917 total segment of data, 158 segments were discarded in this way. 
Of the 433 segments identified at the end of Section \ref{sec:lfqpo}, 428 segments were kept and 5 segments were discarded.
Here we used the measured interquartile range instead of the rms deviation to be more robust against bias from outliers.
This filtering method also has the effect of removing data taken during passage through the polar horns, when there tends to be high particle radiation (as noted in \citealt{Ludlametal18}).

Due to the strong extinction seen from MAXI~J1535--571, there are few intrinsically low-energy source counts below $\approx$\,1\,keV, and the spectrum there is dominated by photons redistributed from higher energies.  
Thus, 1\,keV was chosen as the lower energy bound for the soft band in the analysis.
At high energies, we found that the background was greater than or equal to the source counts above $\approx$\,10\,keV, so we used 10\,keV as the higher energy bound for the hard band in the analysis.
The broad energy band used for power spectral analysis and the lag reference band is 3--10\,keV so that the results can be reasonably compared to previous \emph{RXTE} results, and because LF QPOs are typically stronger at higher energies. 
While the \emph{RXTE} Proportional Counter Array (PCA) had a lower limit of 2\,keV, its effective area varied dramatically in the 2--3\,keV band: the \emph{NICER} XTI is $\approx$\,11$\times$ greater in effective area than the PCA at 2\,keV and is about equal in effective area compared to the PCA at $\approx$\,2.7\,keV. 
Because the effective area curve weights the energy distribution of the average photon counts, we chose to use 3--10\,keV for the broadband analysis.

\section{Results}
\label{sec:results}
\subsection{Source Evolution}

\begin{figure*}
\centering
\includegraphics[width=0.95\linewidth]{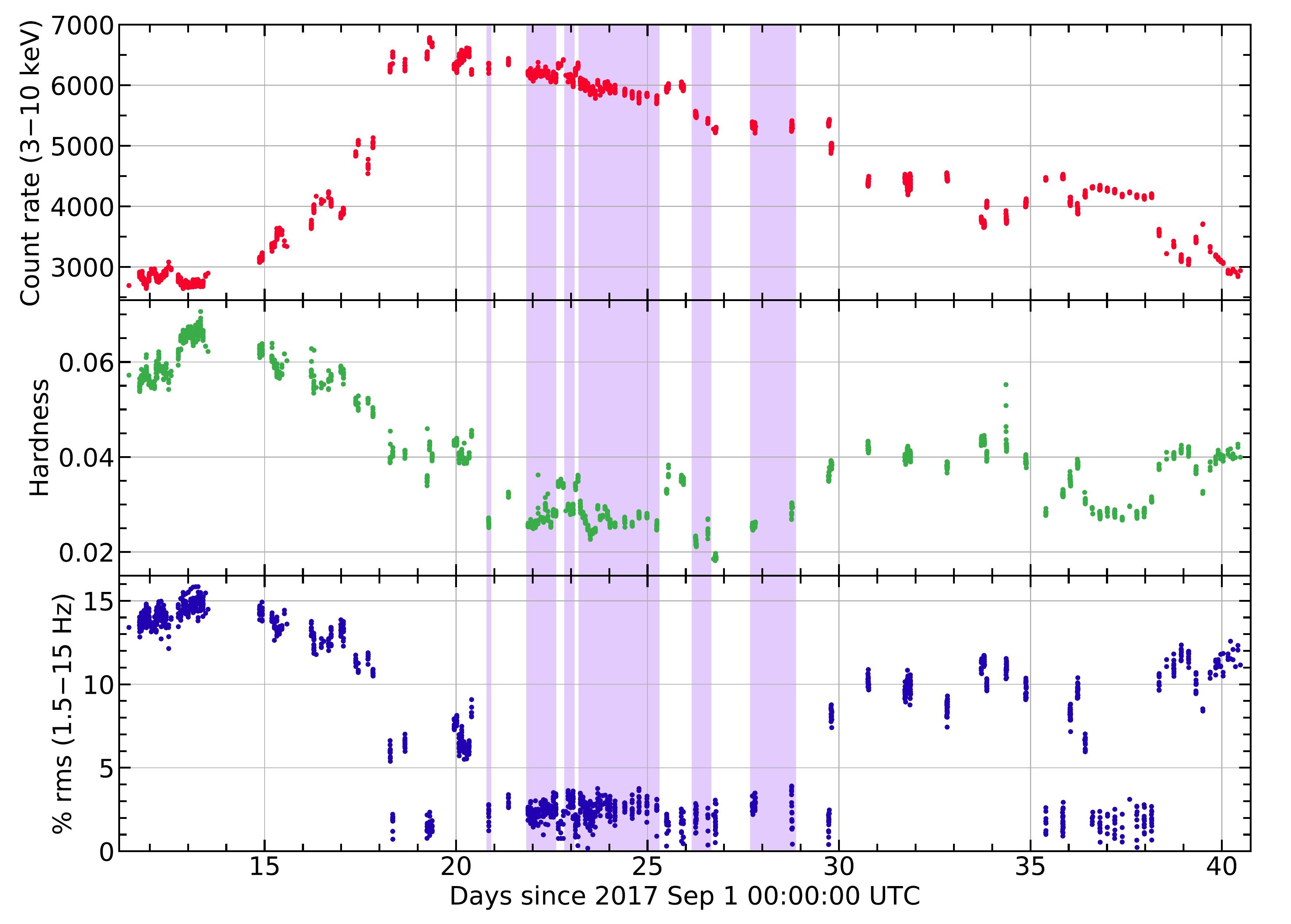}
\caption{
MAXI~J1535--571 spectral and timing properties, for continuous 64\,s segments of the data in the early part of the outburst. 
Time intervals containing segments used in our analysis are marked with purple shaded areas (see Section \ref{sec:lfqpo} and Figures \ref{fig:psdcompare} and \ref{fig:psd-evol} for the selection criteria).
Top panel: average 3--10\,keV count rate (count\,s$^{-1}$).  
Middle panel: spectral hardness ratio (ratio of the count rates in 7--10\,keV to 1--2\,keV bands). 
Bottom panel: fractional rms in 3--10\,keV power spectra, integrated from 1.5--15\,Hz.
}
\label{fig:time-evol}
\end{figure*}

In Figure~\ref{fig:time-evol} we show the 3--10~keV count rate, spectral hardness ratio, and fractional rms evolution of MAXI~J1535--571 in the early part of its outburst for sequential 64\,s segments of the light curve.
Detailed study of the power spectral evolution throughout the observations showed that in the periods with high rms ($\gtrsim$\,5\%) and a higher spectral hardness ratio, the power spectrum is dominated by a strong Type~C QPO with accompanying broadband noise and a strong harmonic (P.!Uttley et al. 2018, in preparation).  
Therefore, we identify those data with the hard-intermediate state (see, e.g., \citealt{Nowak95}; \citealt{Homanetal01}; \citealt{Bellonietal05}). 
Here we focus on the observations showing low rms ($<$\,5\%) and relatively soft spectra, which do not show strong variability, but still show larger fractional variability amplitudes than the canonical soft states observed in other transient black hole X-ray binaries (rms $\lesssim$3\% in 2--20\,keV; \citealt{Bellonietal05,Heiletal15a}). 
We therefore identify these observations with the soft-intermediate state.
The 5\% rms limit was chosen for our analysis due to the gap in rms between 4\% and 5.5\% that is visible in the lower panel of Figure \ref{fig:psdcompare}.
The frequency range 1.5--15\,Hz was used for the rms computation to encompass some of the broadband noise and the expected LF QPO frequency range without being potentially contaminated by very-low-frequency noise or deadtime effects on the high-frequency Poisson noise.
Luckily, as we mention in Section \ref{sec:discuss}, we found that deadtime effects are very minor thanks to \emph{NICER}'s modular design and high throughput.

\subsection{Identifying the LF QPO}
\label{sec:lfqpo}
\begin{figure}
\centering
\includegraphics[width=0.99\columnwidth]{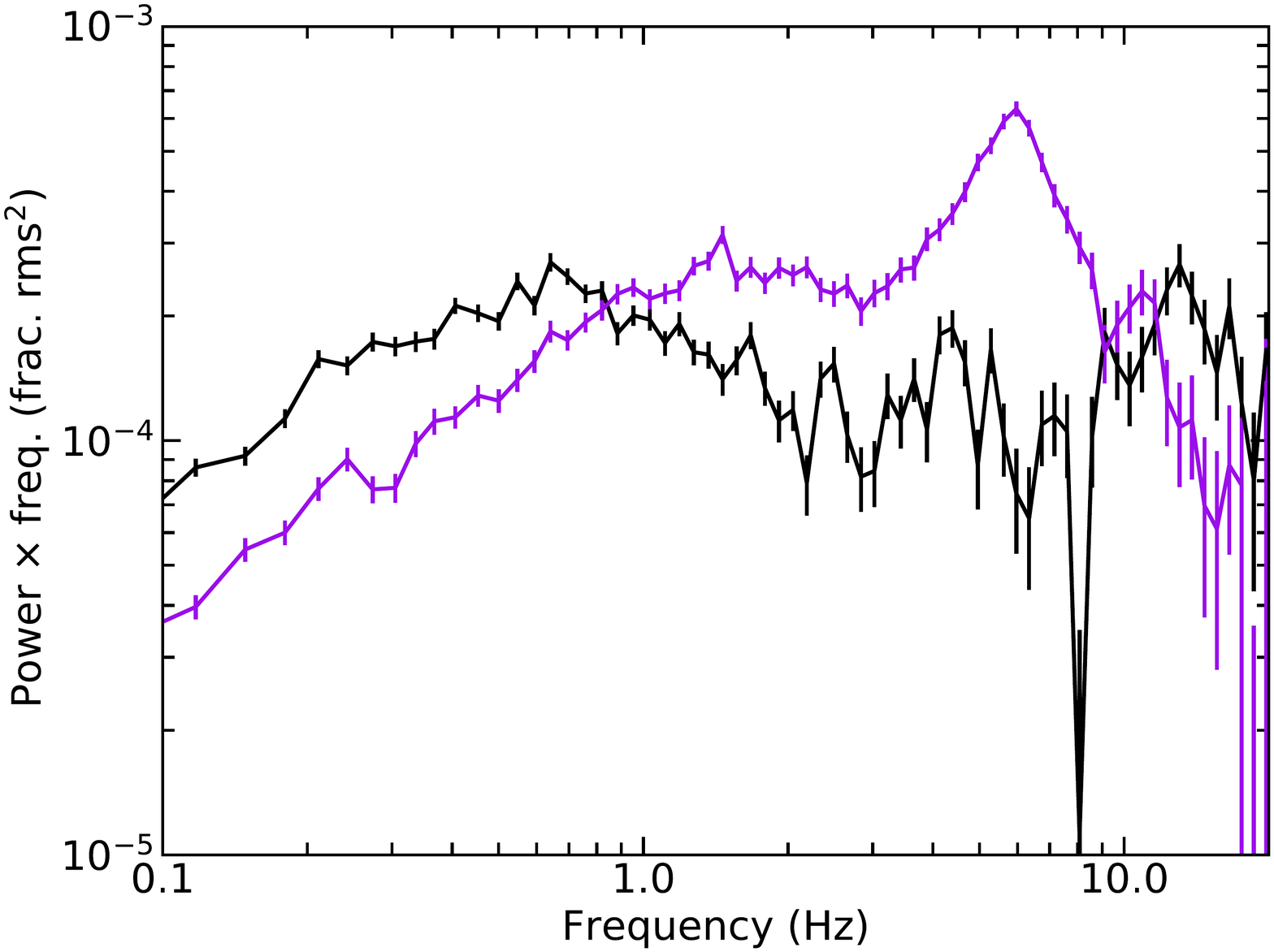}
\includegraphics[width=0.99\columnwidth]{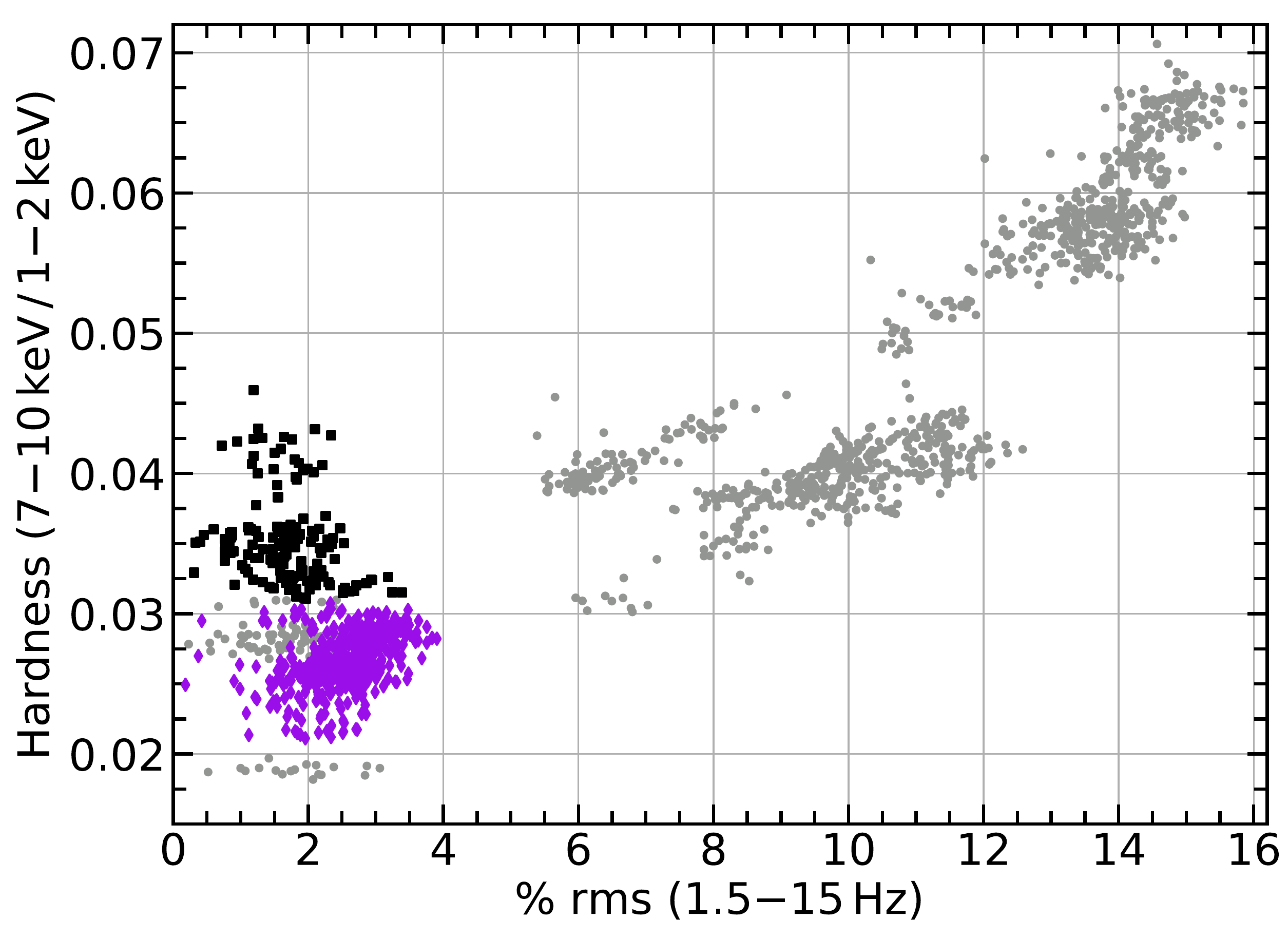}
\caption{
Top: Poisson-noise-subtracted power spectra in 3--10\,keV  averaged from 64\,s segments of data with 1.5--15\,Hz rms\,$<$\,5\%, for two hardness ratio ranges: 0.031--0.050 (black) and 0.021--0.031 (purple). 
A broad QPO-like signal at $\approx$\,5.7\,Hz emerges for the softer hardness ratio range.
The error bars show the 1$\sigma$ level uncertainties.
Bottom: the rms plotted with spectral hardness ratio using the same segment data as in Figure \ref{fig:time-evol}. 
The segments marked with purple diamonds and black squares represent the segments used in the purple and black averaged power spectra in the upper plot. 
The gray dots represent the rest of the segments from Figure \ref{fig:time-evol}. }
\label{fig:psdcompare}
\end{figure}

We looked for Type~A and B QPOs in the soft-intermediate state observations of this source.
These QPOs stand out above the broadband noise in the averaged power spectrum and their frequency tends to stay constant in time for a given flux, which lets us stack segments easily. 
However, they are only present in a small hardness window within the soft-intermediate spectral state \citep{Mottaetal11}, and Type B QPOs sometimes appear to sharply turn off and on as a function of time \citep{Bellonietal05}.
Because QPOs cannot be readily identified in 64\,s sections of the weakly-varying soft-intermediate state light curves, we searched for evidence of characteristic timing signatures by sorting the data according to the spectral hardness ratio (the ratio of the 7--10\,keV to 1--2\,keV count rate). 
In data from other X-ray binaries, Type A and B QPOs have been known to appear only in narrow ranges of hardness as the system evolves during an outburst (e.g., \citealt{Belloni10}).
We then selected 64\,s segments in a range of hardness ratios and computed the 3--10\,keV average power spectrum; this step was repeated sliding the hardness ratio window by 30\% increments.
This analysis revealed the presence of a fairly broad QPO-like signal at a frequency of $\approx$\,5.7\,Hz, which is present for hardness ratios $\leq$\,0.031 but not for harder power spectra (plotted with Poisson noise subtracted in the upper panel of Figure~\ref{fig:psdcompare}). 

\begin{figure}
\centering
\includegraphics[width=0.99\columnwidth]{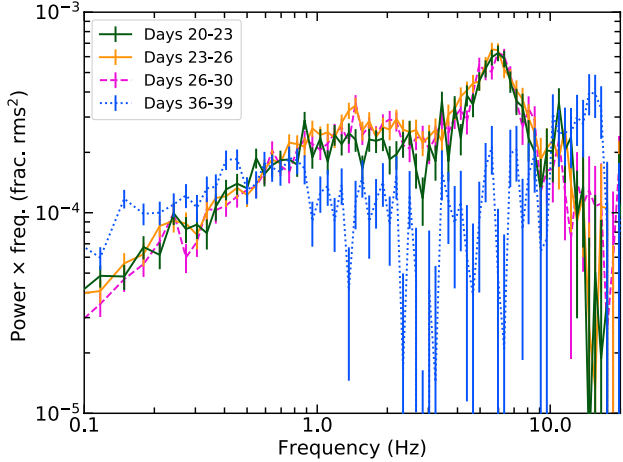}
\caption{
Poisson-noise-subtracted power spectra in the hardness ratio range 0.021$-$0.031 in four time windows: days 20$-$23, 23$-$26, 26$-$30, and 35$-$39. 
Day 0 is 2017 September 1, as in Figure \ref{fig:time-evol}.
The $\approx$\,5.7\,Hz QPO-like feature is visible in the power spectra in the first three time windows, but not in the fourth window.
The error bars show the 1$\sigma$ level uncertainties.
}
\label{fig:psd-evol}
\end{figure}

The presence of an $\approx$\,5.7\,Hz QPO-like feature can be excluded at a $3\sigma$ upper limit of $4.480\times10^{-5}$\,Hz$^{-1}$ in normalization (0.64\% in rms) in the averaged power spectrum of the harder segments (see Figure \ref{fig:psdcompare}).
The QPO-like peak seems to shift in frequency and weaken significantly for the lowest hardness ratios ($<$\,0.02), but appears to be relatively stable in frequency over the hardness ratio range 0.021--0.031. 
We also checked the power spectra in that same hardness range in four different count rate windows that correspond to days 20$-$23, 23$-$26, 26$-$30, and 35$-$39 (with 2017 September 1 as Day 0, as in Figure \ref{fig:time-evol}) to assess if the count rate or elapsed time in outburst affected the variability. 
The Poisson-noise-subtracted power spectra for these four time windows are plotted in Figure \ref{fig:psd-evol}. 
The $\approx$\,5.7\,Hz QPO-like feature is evident in the first three windows but not in the last window; there, its $3\sigma$ upper limit in normalization is $1.080\times10^{-4}$\,Hz$^{-1}$ (1.0\% in rms) and there is a stronger possible feature at $\sim$\,12\,Hz.
Therefore, for the rest of our study we combined the data from 64\,s sections, selected in the hardness ratio range 0.021--0.031 with an integrated 1.5--15\,Hz rms\,$<$\,5\% and a 3--10\,keV count rate greater than 5000 counts\,s$^{-1}$. 
We note that the QPO-like feature in Figure \ref{fig:psd-evol} appears to be intrinsically broad and stable in frequency for the first three time windows.


There are 428 good 64\,s continuous sections of data that fit our spectral hardness, rms, and count rate criteria, which gives a total exposure of 27.392\,ks. 
The average count rate in the 3--10\,keV energy band is 5921\,count~s$^{-1}$.
For the reference energy band used in Section \ref{sec:lags}, which also covers 3--10\,keV but uses MPUs 4--6, the average count rate is 2692\,count~s$^{-1}$. 

\subsection{Average Power Spectrum} \label{sec:power}

\begin{figure}
\centering
\includegraphics[width=0.99\columnwidth, clip]{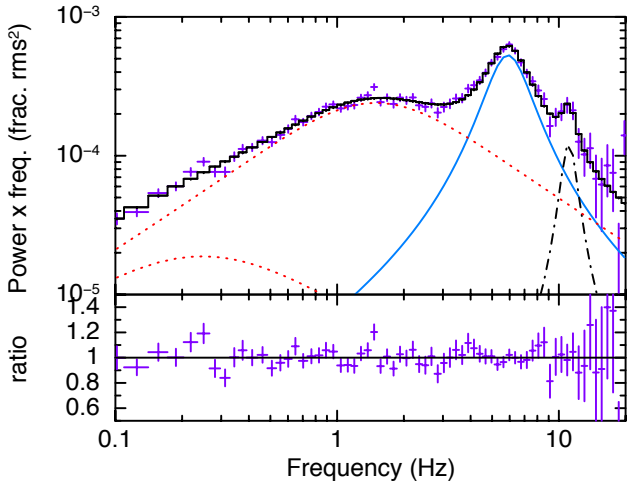}
\caption{
Power spectrum 
in the energy band 3--10\,keV with the Poisson noise subtracted, same as the purple power spectrum in Figure \ref{fig:psdcompare}.
The data (purple points), total model (solid black), and Lorentzian models for the QPO (solid blue), QPO harmonic (dash-dotted black), and broadband noise (dotted red) are plotted in the upper panel.
The lower panel shows a ratio of the total model to the data.
}
\label{fig:powerspectra}
\end{figure}

We computed average power spectra (Figure \ref{fig:powerspectra}) in the energy band 3--10\,keV for the 428 good segments described above, with all XTI MPUs, using our own code\footnote{The analysis software and processed data products will be available in the GitHub repository 
\url{https://github.com/abigailStev/MAXIJ1535\_QPO} after the publication of this Letter.} (see \citealt{vdKlis89} for an overview of Fourier-domain methods).
The power spectra were geometrically re-binned in increments of 6\% of the frequency.  
We fitted the power spectra over the range 0.03--128\,Hz with four Lorentzians: two model the broadband noise, one the QPO fundamental, and one the weak QPO harmonic. 
From these fits, we found the \vcent, FWHM, and normalization of the Lorentzian components.
The Poisson noise was modeled with a power-law with a slope of 1 for fitting $\nu P_{\nu}$.
The best-fitting parameter values are listed in Table \ref{tab:pow}. 
The errors represent the 90\% confidence region computed from the \textsc{xspec} MCMC routine.

The QPO-like feature has a centroid frequency of $\vcent=5.72^{+0.04}_{-0.06}$\,Hz.  
It is relatively broad, and the fitted width is influenced by the bump on the left of the peak at $\sim$\,4\,Hz and the scatter on the harmonic at $\gtrsim$\,9\,Hz.
Thus, we heuristically determined and fixed the quality factor at $Q=2$. 
The corresponding FWHM of the feature is 2.86\,Hz. \footnote{If the FWHM of the QPO is a free-fitting parameter, we find $\text{FWHM}=3.0\pm0.2$\,Hz, giving $Q=1.9$.}
The QPO-like feature contributes a 3--10\,keV rms of 1.9\% and a 5--10\,keV rms of 2.9\%.
Although this feature is quite broad compared to Type C QPOs, it is distinct from the broadband noise and possesses a clear --- if weak --- harmonic.  
Therefore, we will refer to it as a QPO for the remainder of this Letter.
The closest analogy to this feature appears to be the QPOs originally identified by \citet{Wijnandsetal99} and \citet{Homanetal01} as Type~A-I and Type~A-II QPOs in the black hole transient XTE~J1550--564; Type~A-I QPOs were later identified by \citet{Casellaetal05} as a form of Type~B QPO.

The Type~A-I (later Type~B) QPO in XTE~J1550--564 has a frequency of 5.9\,Hz, a $Q$-value of $Q=2.4$, 3\% rms in 2--60\,keV, a harmonic at 11\,Hz, and it was present in the soft-intermediate state. 
The Type~A-II QPO in XTE~J1550--564 has a frequency of 8.5\,Hz, a $Q$-value of $Q=2.2$, 2\% rms in 2--60\,keV, no harmonic, and it was also present in the soft-intermediate state. 
This MAXI~J1535--571 QPO has a similar fundamental frequency and harmonic frequency to the Type~A-I, but Q-value and rms more similar to the Type~A-II (though it may not be inconsistent with the Type A-I when considering differences in the energy-dependent effective area of the \emph{RXTE} PCA and \emph{NICER}). 
Based on these shared characteristics, the QPO could be a Type A or Type B, so we refer to it as a Type~A/B. 

\begin{table}  
\centering
\begin{tabular}{l l l l}
\toprule
Component & Parameter & Value & Notes\\
\midrule
Poisson noise & norm.\,($\times10^{-4}$\,Hz$^{-1}$) & $3.357\pm 0.003$ & \\
BBN$_{1}$ & FWHM~(Hz) & $0.3^{+0.2}_{-0.02}$ &  \\
BBN$_{1}$ & \vcent~(Hz) & $1\times10^{-22}$ & [1] \\
BBN$_{1}$ & norm.\,($\times10^{-4}$\,Hz$^{-1}$) & $8^{+5}_{-1}$ & \\
BBN$_{2}$ & FWHM~(Hz) & $2.8^{+0.1}_{-0.2}$ &  \\
BBN$_{2}$ & \vcent~(Hz) & $0.5^{+0.2}_{-0.1}$ & \\
BBN$_{2}$ & norm.\,($\times10^{-3}$\,Hz$^{-1}$) & $1.1\pm 0.1$  & \\
QPO & FWHM~(Hz) & 2.86 & [2] \\
QPO & \vcent~(Hz) & $5.72^{+0.04}_{-0.06}$ &  \\
QPO & norm.\,($\times10^{-4}$\,Hz$^{-1}$) & $4.1\pm 0.2$ & \\
Harmonic & FWHM~(Hz) & $2^{+2}_{-0.5}$ & \\
Harmonic & \vcent~(Hz) & $11.1^{+0.3}_{-0.4}$ & \\
Harmonic & norm.\,($\times10^{-5}$\,Hz$^{-1}$) & $3^{+2}_{-1}$ & \\
\bottomrule
\end{tabular}
\caption{ 
Best-fitting parameter values for the power spectrum with Poisson noise, with $\chisq=102.25$ for 93 Degrees of Freedom. 
The errors on the parameters represent the 90\% confidence region.
Final column notes: 
[1]: Frozen at $1\times10^{-22}$ to represent the very-low-frequency broadband noise and eliminate degeneracy between the two broadband noise components.
[2]: Fixed at half the QPO centroid value.
\label{tab:pow}}
\end{table}

\subsection{Lag-energy Spectrum} \label{sec:lags}

\begin{figure}
\centering
\includegraphics[width=0.99\linewidth]{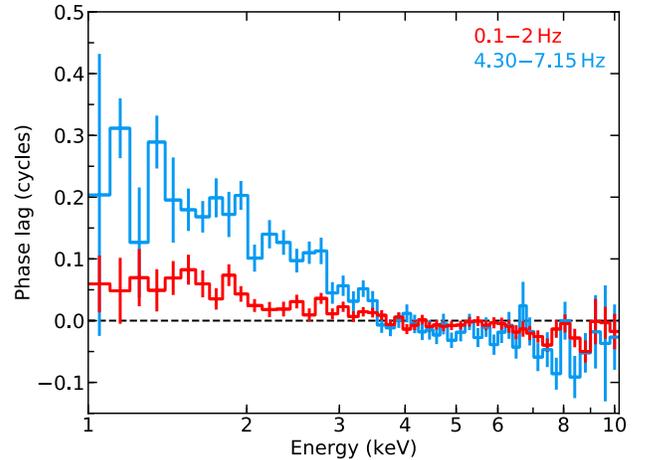}
\caption{
Lag-energy spectra computed over the frequency ranges 0.1--2.0\,Hz (red: broadband noise) and 4.29--7.15\,Hz (blue: QPO) plotted in phase normalized to 1. 
One ``cycle'' refers to the characteristic timescale of the variability, that is, 0.175\,s for the QPO quasi-period or 1\,s for the middle of the frequency range used for the broadband noise.
We see soft lags for the broadband noise and QPO.
\label{fig:lag-energy}}
\end{figure}

The lag-energy spectrum, computed from the average cross spectrum, measures by how much the variability in many narrow energy bands leads or lags the variability in a broad `reference' energy band (see \citealt{Uttleyetal14} for an overview of lag-energy spectra with examples and recipes).
To ensure that the narrow-band and broad reference band light curves are truly independent, we use the detector information stored in the event lists as a selection criterion.
We used MPUs 0--3 (inclusive) when extracting narrow bands and MPUs 4--6 (inclusive) for extracting the reference band. 
This division of MPUs was chosen to optimize the signal-to-noise in the narrow bands while maintaining enough signal in the reference band.
As noted in Section \ref{sec:data}, we removed FPMs 11, 14, 20, 22, 34, and 60, so that there are 50 total FPMs: 27 were used for the channels of interest, and 23 for the reference band.
To further improve signal-to-noise, we binned coarser than the intrinsic energy resolution of the data and achieved 50 narrow-band `channels of interest' across the energy range 1--10\,keV.
The channels of interest are approximately equally spaced on a log scale (see Figure \ref{fig:lag-energy}). 

To compare the lag-energy spectra of the broadband noise and the QPO components, cross-spectra were averaged and resulting phase and time-lags calculated across three integrated frequency ranges: 0.1--2.0\,Hz (intrinsic broadband noise), 4.29--7.15\,Hz (QPO), and 9.6--12.6\,Hz (harmonic).  
The frequency ranges used for the QPO and harmonic were chosen to be the FWHM centered on the centroid for the fitted Lorentzian model in the power spectrum (Table \ref{tab:pow}; Figure \ref{fig:powerspectra}). 
The lag-energy spectra for the broadband noise and QPO are shown in Figure \ref{fig:lag-energy}.

The broadband noise lags show a ``soft lag,'' in which the variability at softer energies lags behind the variability at harder energies.  
The QPO lags also show a definitive soft lag shape, with a flattening above 4\,keV.  

The shape of the QPO lag-energy spectrum is reminiscent of the inflected lag-energy spectrum shape (like a broken log-linear shape) observed for the Type~B QPO of GX~339--4 \citep{StevensUttley16}, but with the {\it opposite} sign of lags.  
We note for completeness that we also computed the lags over the same frequency range of the segments that were excluded due to their hardness values, as a check that this soft lag is due to the presence of the QPO. 
While it is not possible to say much about lag features due to signal-to-noise limitations, the lag-energy spectrum does not appear to follow the same trends.
The harmonic lag-energy spectrum, computed for the frequency range 9.6$-$12.6\,Hz, has better signal-to-noise than the power spectrum of the harder segments (black points in Figure \ref{fig:psdcompare}), but still has larger error bars than the QPO lag-energy spectrum.  
There is significant scatter and it does not show any trends or features.

\section{Discussion and Conclusion} \label{sec:discuss}

We have reported the discovery and initial spectral-timing analysis of a QPO feature in the soft-intermediate state of the 2017--2018 outburst of the new candidate black hole transient MAXI~J1535--571.  
This QPO has a centroid frequency of $5.72^{+0.04}_{-0.06}$\,Hz and is relatively broad (Q\,$\approx$\,2). 
It is also weak, with an integrated fractional rms of 1.9\% in 3--10\,keV and 2.9\% in 5--10\,keV.
The QPO shows large soft lags with respect to the 3--10\,keV reference band of about 29\% of a QPO cycle at 1.3\,keV ($\sim$\,50\,ms), and the broadband noise shows 5\% phase lags at 1.3\,keV (also $\sim$\,50\,ms).
The broadband noise shows a smoother lag-energy spectrum compared to the inflected shape of the QPO lag-energy spectrum.
The QPO lags noticeably flatten above $\sim$\,4\,keV. 

Soft lags (i.e., positive at softer energies) are also observed for the similar QPOs in XTE~J1550--564, albeit across the harder energy range 2--48\,keV \citep{Wijnandsetal99}. 
The lags that we see are approximately of the same magnitude and the same sign as the lags in \citet{Wijnandsetal99} when comparing the overlapping energy range of \emph{NICER} and \emph{RXTE}.

The QPO analyzed in this Letter has some notable differences with the clear Type~B QPO in GX~339--4 analyzed with lag-energy spectra and phase-resolved spectroscopy in \citet{StevensUttley16}.
There, the QPO had a fractional rms of $\sim$\,14\% in 5--10\,keV, and the broadband noise was much weaker relative to the QPO, but larger in normalization than we see here. 
The QPO also showed \emph{hard} lags of about 10\,ms ($\approx$\,30\% in normalized phase).   
It is notable that the inflected shapes of the lag-energy spectra observed in MAXI~J1535--571 and GX~339--4 are similar, but with opposite sign.  
The inflection may be due to the phase offset between sinusoidally varying spectral components, as shown in simulations in \citet{StevensUttley16} and \citet{Stevens18}.
Thus the inflected shape of the lag-energy spectrum in MAXI~J1535--571 would then be linked to a phase offset between the peaks in power-law emission and the modulation of the disk blackbody spectrum, where the disk component would \emph{lag} rather than lead the power-law emission. 
In GX~339--4 we inferred that there was a large 30\% phase shift from spectral fitting of the Wien tail, while with \emph{NICER} data of MAXI~J1535--571 we can probe low-enough energies to measure this lag directly.

The proposed mechanism for the Type B QPO in GX~339--4 was a large-scale-height precessing power-law emitting region such as the base of the jet \citep{StevensUttley16}.  
There, the phase lead of the disk relative to power-law variations in GX~339--4 was attributed to a geometry where the approaching side of the disk is illuminated by the jet base, leading to enhanced (blue-shifted) disk emission, before the jet base points toward the observer leading to enhanced power-law emission (e.g., due to beaming effects, optical depth effects, and other aspects of the jet emission mechanism). 
GX~339--4 is probably a low-inclination system,\footnote{The binary inclination angle of GX~339--4 has been constrained to $37\degrees < i \lesssim 60\degrees$ from X-ray \citep{Zdziarskietal98} and optical \citep{Heidaetal17} observations. Recent spectral analysis by \citet{WangJietal18} estimates $i\approx 40\degrees$ from spectral modeling.}
while MAXI~J1535--571 may be a high-inclination system (suggested by spectral fits of the iron line in the hard-intermediate state; \citealt{Xuetal18,Milleretal18}).
The opposite lag sign in MAXI~J1535--571 could then suggest that the maximum in power-law emission is seen when the jet is pointing away from the observer (e.g., if solid angle effects dominate over beaming effects for the expected larger offset of the jet axis to the line of sight).  

We also note that XTE~J1550--564, which shows LF QPOs with similar properties to those seen in MAXI~J1535--571, is a high-inclination (more edge-on) source ($74\degrees.7\pm3\degrees.8$; \citealt{Oroszetal11}) with the jet and binary axes in alignment \citep{Steineretal12}. 
However, XTE~J1550--564 also shows ``normal'' Type~B QPOs \citep{Homanetal01} that show the same lag sign as in GX~339--4 (i.e., opposite to what the same source shows for the different Type~A QPOs; \citealt{Remillardetal02}).
This suggests that the sign of the lag may instead (or also) be due to the scale height of the emitting region, which then affects whether optical depth effects are important when considering whether the power-law flux variation lags or leads the soft flux variation. 
The changing sign of the lag in the same source is then linked to a change in the geometry (scale height) of the emitting region (as it is not possible for the inclination of a source to rapidly change). 
The sign of the lag in MAXI~J1535--571, and the \emph{time} lag being $\sim$\,50\,ms for both the QPO and the broadband noise, could therefore mean that there is a compact, self-interacting emitting region that strongly couples the broadband noise lag to the QPO lag. 
Our hypothesis on the emission region and mechanism for this QPO will be rigorously tested in a follow-up paper using phase-resolved spectroscopy analysis as in \citet{StevensUttley16} and \citet{Stevens18}.
It is possible that MAXI~J1535--571 also exhibited a ``normal'' Type B QPO with a higher $Q$-value and higher rms, and it was missed by \emph{NICER} due to a lengthy Sun-avoidance data gap in winter 2017--2018.

It is interesting to note that the fitted Poisson noise level in Section \ref{sec:power} is only 1\% lower than that expected from the mean count rate (noise level of $2/\text{count rate} = 2/5921 = 3.378\times 10^{-4}$ in units of fractional rms$^{2}$~Hz$^{-1}$). 
Poisson noise levels are reduced by instrumental deadtime effects, which suppress the observed noise variance due to the resulting (anti-)correlations between successive photon counts.  
The data show that the fraction of photons lost to deadtime in the \emph{NICER} detectors is indeed remarkably small for such a bright source ($\sim$\,16,000\,counts\,s$^{-1}$ for the whole bandpass).

Finally, we note that these observations highlight the enormous potential of \emph{NICER} for transforming our understanding of accreting compact objects via spectral-timing methods.  
At the peak flux levels of the soft-intermediate state of MAXI~J1535--571, the total \emph{NICER} count rate exceeded 16,000 count\,s$^{-1}$, with very little deadtime and hence minimal spectral distortion.  
Furthermore, these large phase lags would not have been observable by the \emph{RXTE} PCA due to the cut off in response below $\approx$\,3\,keV.
The combination of such large count rates with good energy resolution and a soft X-ray response is a revolutionary capability.
\emph{NICER} points the way to a bright future for our understanding of the innermost regions of accreting compact objects.

\acknowledgments
The authors appreciate the \emph{NICER} data reduction pipeline development led by C.B.~Markwardt.
A.L.S. is supported by an NSF Astronomy and Astrophysics Postdoctoral Fellowship under award AST-1801792.
D.A. and A.I. acknowledge support from the Royal Society.
E.M.C. gratefully acknowledges support from NSF CAREER award AST-1351222.
A.C.F. acknowledges support from ERC Advanced Grant 340442.
R.M.L. acknowledges funding through a NASA Earth and Space Science Fellowship.
D.R.P. is supported by NASA through an Einstein fellowship (PF6-170156).
J.F.S. has been supported by NASA Einstein Fellowship grant No. PF5-160144.  
J.v.d.E. acknowledges support from NWO.
This work was supported by NASA through the \emph{NICER} mission and the Astrophysics Explorers Program. 
This research made use of data and software provided by the High Energy Astrophysics Science Archive Research Center (HEASARC) 
and NASA's Astrophysics Data System Bibliographic Services.

\vspace{5mm}
\facilities{\emph{NICER}, ADS, HEASARC}

\software{Astropy v2.0.3 \citep{AstropyPaper}, HEASoft v6.24, Jupyter notebooks v4.4.0, Matplotlib v2.1.2 \citep{Matplotlib}, NumPy v1.14.0, Python v3.6.4, SciPy v1.0.0, XSPEC v12.10.0c \citep{XSPEC}}



\end{document}